\documentstyle[art12]{article}
\begin{document}
\title{{\bf Generalized Exclusion Statistics in the Kondo Problem}}
\author{{\large A. P. Protogenov$^{(1)}$ and V. A. Verbus$^{(2)}$}}
\maketitle
\begin{center}
{\footnotesize {\em
$^{(1)}$Institute of Applied Physics, Russian Academy of Sciences,\\
46 Ul'janov Street, Nizhny Novgorod 603600, Russia \\
$^{(2)}$Institute for Physics of Microstructures, Russian Academy of Sciences,\\
46 Ul'janov Street GSP-105, Nizhny Novgorod 603600, Russia}}
\end{center}

\begin{abstract}
We consider the generalized exclusion 
statistics in the Kondo problem. The thermodynamic 
Bethe ansatz equations have been used for a multicomponent 
system of particles obeying the 
generalized exclusion principle. 
We have found a relation between the derivative of the phase 
shift of the scattering matrix for Fermi particles and 
for particles characterized by generalized exclusion statistics. 
We show that the statistical matrix in the Kondo problem has 
a universal form in high and low temperature limits.
\end{abstract}
%
%
%
%
\section*{1. Introduction}
\par
Generalized exclusion statistics (GES) \cite{hal} has attracted much
interest in recent papers [2 -- 8] 
A typical example of
the application of generalized exclusion statistics is the
Cologero-Sutherland model \cite{cal,suth,ha}. D. Bernard and
Y.-S. Wu \cite{ber} studied the GES in the thermodynamic Bethe
ansatz (TBA) using equivalence of the
Bethe ansatz (BA) equations and the ones obtained by the Haldane
principle \cite{ha}. M. Wadati has shown  \cite{wad} that a
change in the statistics is determined by an appropriate choice of the
phase shift of the scattering matrix. The generalization of 
exclusion statistics for multicomponent 
systems was provide in the paper \cite{fu}.
\par
The goal of this paper is to consider the exclusion statistics
of excitations in the Kondo problem. To solve this problem we
have applied the TBA for a multicomponent system of particles obeying
the GES. We have found the universal
behavior of the statistical matrix in high- and low-temperature
limits. In these limits the statistical matrix is proportional to
the Cartan matrix for the $A_{n}$-algebra.
\par
The paper is organized as follows. In the second section we provide 
the main equations of the exclusion statistics theory to complete the
discussion. In this section we have also derived the
relation between the derivative of phase shift (DPS) of the 
scattering matrix and
the statistical matrix in the framework of the TBA equations
for the multicomponent system of particles. The third section 
deals with the application of this result to the Kondo
problem. We show that distribution function and statistical
matrix in this problem have the universal form in high and
low temperature limits.
%
%
%
%

\section*{2. Exclusion Statistics Equations}
\par
A change in the number $D_{\alpha i}$ of the vacant states
due to the addition of the number  $N_{\beta j}$ of the particles,
according to Haldane \cite{hal} is defined as
\begin{equation}
\frac{\partial D_{\alpha i}}{\partial N_{\beta j}} 
= -g_{\alpha i,\beta j}\, .
\label{E1}
\end{equation}
Here  $g_{\alpha i,\beta j}$ is the matrix of statistical interaction.
The indices $\alpha$ $(\alpha = 1,2,...M)$ and $i$ correspond to the
internal and the dynamical degrees of freedom, respectively.
The solution of Eq.(\ref{E1}) has the form:
\begin{equation}
D_{\alpha i} = -\sum_{\beta j} g_{\alpha i,\beta j}
N_{\beta j} + D^{0}_{\alpha i}\, ,
\label{E2}
\end{equation}
where $D^{0}_{\alpha i}$ is the number of vacant states of the
$\alpha i$-th type without particles. The number of holes 
$D_{\alpha i}$ determines the statistical weights $W$ as follows
\begin{equation}
W = \prod_{\alpha,i}\frac{(N_{\alpha i}-1+D_{\alpha i}(\{N_{\beta j}\})
+\sum_{\beta j} g_{\alpha i,\beta j}\delta_{\alpha\beta}\delta_{ij})!}
{(N_{\alpha i})!(D_{\alpha i}(\{N_{\beta j}\})-1+\sum_{\beta j}
g_{\alpha i,\beta j}\delta_{\alpha\beta}\delta_{ij})!}\, .
\label{E3}
\end{equation}
In the specific cases of $g_{\alpha i,\beta j}=0$ and $g_{\alpha i,\beta j}=
\delta_{\alpha\beta}\delta_{ij}$ Eq.(\ref{E3}) yields the well-known
expression for the statistical weights of Bose and Fermi particles.
\par
The distribution function $n_{\alpha i}$ is defined usually
\cite{wu,ber,isak} in the following way
$n_{\alpha i}=N_{\alpha i}/G_{\alpha i}^{0}$.
This is not convenient for systems with internal degrees of freedom.
For example, in the hierarchical basis of the states in the fractional
quantum Hall effect $n_{\alpha i}=\infty $ if
$\alpha = 2,3,...$ because $G_{\alpha i}^{0}=0$ for spin degrees of freedom.
The definition of the distribution function in the form
\begin{equation}
n_{\alpha i} = \frac{N_{\alpha i}}{G_{\alpha i}}
\label{E4}
\end{equation}
with
\begin{equation}
G_{\alpha i} = G^{0}_{\alpha i} +
N_{\alpha i}-\sum_{\beta j} g_{\alpha i,\beta j}N_{\beta j}
\label{E5}
\end{equation}
is more convenient because it takes into account the influence
of the statistical interaction on the number of the states. 
The equilibrium distribution function $ n_{\alpha i}$ can
be found in this case from the extremum of the grand partition function
as a solution of the following equations:
\begin{equation}
\frac{1}{n_{\alpha i}}\prod_{\beta j}\left[ 1-n_{\beta j}\right]^
{g_{\beta j,\alpha i}} =
\exp\{(\epsilon_{\alpha i}-\mu_{\alpha})/T\}\, .
\label{E6}
\end{equation}
Here $\epsilon_{\alpha i}$ is the energy and $\mu_{\alpha}$
is the chemical potential for the particles of type $\alpha$.
The distribution function $n_{\alpha i}$ determines the free
energy
\begin{equation}
F = \sum_{\alpha}\mu_{\alpha}N_{\alpha} - T\sum_{\alpha i}
G^{0}_{\alpha i}\ln \left[ \frac{1}{1+n_{\alpha i}}\right]
\label{E7}
\end{equation}
as well as the value of the entropy $S=\ln W$.
\par
The interection between quasiparticles from the TBA point of view is encoded
in the phases $\Theta_{\alpha i,\beta j}$ of the dynamical scattering matrix
$S_{\alpha i,\beta j}=-\exp(-i\Theta_{\alpha i,\beta j})$.
The statistical properties expressed by the statistical matrix
$g_{\alpha i,\beta j}$ depend on the DPS of the scattering 
matrix \cite{ber,wad}.
\par
Let us consider the TBA equations for the set of multicomponent
particles obeying the GES.
Quantizing a gas of such particles on a
circle of length $L$ requires that the momentum $k_{\alpha i}$
of the $\alpha i$-th particle  satisfies the following condition:
\begin{equation}
\exp \{ik_{\alpha}(\theta_{i})L\}\sum_{\beta j}^{N}S_{\alpha ,\beta }
(\theta_{i}-\theta_{j}) = 1\, .
\label{E8}
\end{equation}
The momentum and the energy of the particles are parametrized by the
rapidity $\theta$. Going to the limits $L\rightarrow\infty$
and $N\rightarrow\infty$ with  the finite value of $N/L$ and
taking the derivative of the $\log$ of Eq.(\ref{E8}) yield
\begin{equation}
2\pi q_{\alpha}(\theta) = \frac{dk_{\alpha }(\theta)}{d\theta} + \sum_
{\beta =1}^{M}\int_{-\infty}^{\infty}\Phi_{\alpha \beta}(\theta-\theta')
\rho_{\beta}(\theta')d\theta'\, ,
\label{E9}
\end{equation}
where
\begin{equation}
\Phi_{\alpha\beta}(\theta)=\frac{1}{i}\frac{d}{d\theta}\ln S_{\alpha\beta}
(\theta)\;
\label{E10}
\end{equation}
is DPS.
The function $\rho_{\beta}(\theta)$ in Eq.(\ref{E9}) is the density of the
particles of $\beta$-type, $q_{\alpha}(\theta)$ is the density of the states.
\par
The information about the statistical properties of the system is
contained  in the distribution functions $n_{\alpha i}$ and in the entropy $S$.
In the framework of the thermodynamic Bethe ansatz \cite{zam}, we
have neither information about the ground state energy, nor about
the structure of low-lying excitations with the energy
$\epsilon_{\alpha}^{0}(\theta)$. They are the objects of the
traditional Bethe ansatz approach.
The equilibrium state at temperature $T$ is obtained by minimizing
the free energy $F=E-TS$, where the energy of the system is
\begin{equation}
E = \sum_{\alpha = 1}^{M}\int\epsilon_{\alpha}^{0}(\theta)\rho_{\alpha}
(\theta)d\theta\, .
\label{E11}
\end{equation}

The variation of $F$ with respect to $\rho_{\alpha}$ yields the
following equations for the dressed energy $\epsilon_{\alpha}(\theta)$:
\begin{equation}
\epsilon_{\alpha}(\theta) = \epsilon_{\alpha}^{0}(\theta) +\mu_{\alpha}
- \frac{T}{2\pi}\sum_{\beta=1}^{M}\int \Phi_{\beta\alpha}(\theta-\theta')
\ln \left[\frac{1}{1-n_{\beta}(\theta')}\right]d\theta'\, .
\label{E12}
\end{equation}
Here the function $n_{\alpha}$ relates to the functions $\rho_{\alpha}$
and $q_{\alpha}$ in Eq.(\ref{E9}) as follows: $n_{\alpha}=\rho_{\alpha}/
q_{\alpha}$.
\par
Let us assume that the particles are fermions, i.e.,
$g_{\beta j,\alpha i}=\delta_{\alpha\beta}\delta_{ij}$. We see from
Eq.(\ref{E6}) that
\begin{equation}
n_{\alpha} =\frac{1}{1+ \exp \left[(\epsilon_{\alpha}^{f}(\theta)
-\mu_{\alpha})/T\right]}\, .
\label{E13}
\end{equation}
After substituting this expression into Eq.(\ref{E12}), we have the standard
TBA equations
\begin{equation}
\epsilon_{\alpha}^{f}(\theta) = \epsilon_{\alpha}^{0}(\theta) +\mu_{\alpha}
- \frac{T}{2\pi}\sum_{\beta=1}^{M}\int\Phi_{\beta\alpha}^{f}(\theta-\theta')
\ln \left[1+\exp \left(-\frac{\epsilon_{\beta}^{f}(\theta')-
\mu_{\beta}}{T}\right)\right]d\theta'
\label{E14}
\end{equation}
for fermions (superscript $"f"$ denotes the Fermi statistics).
\par
Each statistics corresponds to the specific value of DPS
$\Phi_{\alpha\beta}(\theta)$ in Eq.(\ref{E12}). The transition to the
Fermi statistics leads to the new value $\Phi_{\alpha\beta}^{f}(\theta)$
of this function.  Assuming that the function $n_{\beta}(\theta)$
in Eq.(\ref{E12}) coincides with Eq.(\ref{E13}) we can find
from Eqs.(\ref{E12}), (\ref{E14}) the relation between
DPS $\Phi_{\alpha\beta}(\theta)$  for the arbitrary statistics and
for the Fermi statistics. This relation has the form
\begin{equation}
\Phi(\theta-\theta') = \Phi^{f}(\theta-\theta')
-2\pi\delta(\theta-\theta')[\cal I-\cal G]\, ,
\label{E15}
\end{equation}
where $\cal I$ is the unit matrix and
\begin{equation}
\cal G =
\left(
\matrix {g_{11}     & g_{12}      & g_{13}  & ...  \cr
         g_{21}     & g_{22}      & g_{23}  & ...  \cr
         g_{31}     & g_{32}      & g_{33}  & ...  \cr
         \vdots     &\vdots       & \vdots  & \ddots  \cr}
        \right) \, .
\label{E16}
\end{equation}
is the statistical matrix.
It is easy to see from Eq.(\ref{E15}) that for models with
the function $\Phi_{s\alpha}^{f}(\theta-\theta')\sim \delta (\theta-\theta')$
one can find such statistical matrix $\cal G$ which gives a zero
value of the r.h.s. in this equation.
In these models the phase of the scattering matrix has the structure
of the step function. 
From the TBA point of view these systems look like
a gas of non-interacting particles having the GES.
This case is known \cite{ber,mur} as the ideal exclusion
statistics.
In these models the correlations between particles 
can be transformed to the statistical interaction.
The distribution function of excitations in the systems with
the ideal exclusion statistics can be obtained from Eq.(\ref{E6})
where the statistical matrix is now
$g_{\alpha,\beta}=\delta_{\alpha,\beta}-\frac{1}{2\pi}\Phi_{\alpha\beta}^{f}$
and the dressed energy of excitations coincides with bare energy
$\epsilon_{\alpha}^{0}$. Note that the structure of Eq.(\ref{E6})
looks like that for Eq.(\ref{E15}) (after some simple transformation)
because the function $n_{\alpha}$ satisfies Eq.(\ref{E13}).
%
%
%
%
\section*{2. Generalized exclusion statistics in the Kondo problem.}
\par
Let us consider the Kondo problem from the GES point of view.
Bethe-ansatz equations for the isotropic $s-d$ exchange model
\cite{tz3} in the Kondo problem are
\begin{equation}
\exp (ik_{j}L) = \exp \left(\frac{iIS}{2}\right)\prod_{\gamma}^{P}
\left(\frac{\lambda_{\gamma}+i/2}{\lambda_{\gamma}-i/2}\right)\, ,
\label{E17}
\end{equation}
\begin{equation}
\left(\frac{\lambda_{\gamma}+i/2}{\lambda_{\gamma}-i/2}\right)^{N}
\left(\frac{\lambda_{\gamma}+1/g+iS}{\lambda_{\gamma}+1/g-iS}\right) =
-\prod_{\beta}^{M}\left(\frac{\lambda_{\gamma}-\lambda_{\beta}+i}
{\lambda_{\gamma}-\lambda_{\beta}-i}\right)\, .
\label{E18}
\end{equation}
These equations solve the problem of diagonalization of the
$s-d$ exchange hamiltonian with the impurity spin $S$ and with
the coupling constant $I$. Here the total number of (up-spin) electrons
is denoted as $N (P)$.
The general solutions of Eqs.(\ref{E17}), (\ref{E18})
have the form of $n$-strings according to the string hypothesis
\cite{yang}. The $n$-string is a set of $n$ solutions
given by
\begin{equation}
\lambda_{\gamma}^{(n,j)} = \lambda_{\gamma}^{n} +
i\left(\frac{n+1}{2}-j\right),\qquad  j=1,...,n\, .
\label{E19}
\end{equation}
Here $\lambda_{\gamma}^{n}$ is the real number and $n$
is the order of the string.
The distribution  of the $n$-type particles (holes) density in the
thermodynamic limit  is $\rho_{n}(\lambda)$  ($\rho_{n}^{(h)}(\lambda)$).
Assuming that the particles obey the Fermi statistics, the TBA equations for
the function $\epsilon_{n}(\lambda)$ have the form
\begin{eqnarray}
\epsilon_{n}(\lambda) &=& -\frac{2\epsilon_{F}}{\pi}\tan^{-1}\exp (\pi
\lambda)\delta_{n1} + \frac{T}{2\pi}\int_{-\infty}^{\infty}
\frac{1}{\cosh (\lambda-\lambda')} \nonumber \\
& &\left[\ln\left(1+e^{\epsilon_{n-1}(\lambda')/T}\right)+
\ln\left(1+e^{\epsilon_{n+1}(\lambda')/T}\right)\right]d\lambda'\, .
\label{E20}
\end{eqnarray}
Here and below the dressed energy is $\epsilon_{n}(\lambda)=T\ln [\rho_{n}^{(h)}
(\lambda)/\rho_{n}(\lambda)]$.
\par
The solutions of these nonlinear integral equations describe
the thermodynamic properties of the $s-d$-model.
The external magnetic field $H$ of the problem enters in the boundary
condition as follows:
\begin{equation}
\lim_{n\rightarrow\infty}\frac{\epsilon_{n}(\lambda)}{n} - H = 0\, .
\label{E21}
\end{equation}
Note that this boundary condition means the condition of compensation
of the internal magnetic field $h~=~\lim\limits_{n\to\infty}
(~\epsilon_{n}~/~n)$ by external magnetic field $H$. This
situation takes place as well for some phase states in $(2~+~1)~D$
systems.
\par
The spin free energy
\begin{eqnarray}
F^{sp} &=&-NT\int_{-\infty}^{\infty}\frac{1}{2\cosh (\pi\lambda)}
\ln\left(1+e^{\epsilon_{1}(\lambda)/T}\right)d\lambda \nonumber \\
& &-T\int_{-\infty}^{\infty}\frac{1}{2\cosh (\pi\lambda+1/g)}
\ln\left(1+e^{\epsilon_{2S}(\lambda)/T}\right)d\lambda
\label{E22}
\end{eqnarray}
is expressed by the functions $\epsilon_{n}(\lambda)$.
The first term in the r.h.s. of Eq.(\ref{E22}) corresponds to the spin
free energy in the absence of impurity. The second term is the
impurity contribution to the free energy.
We will focus on the universal properties of the solutions of
Eqs.(\ref{E20}) in the limits $T\to\infty$ and $T\to 0$.
\par
Comparing BA Eqs.(\ref{E20}) and TBA Eqs.(\ref{E14}) one can
see that the relation between them exists if
\begin{equation}
\epsilon_{n}^{f}(\lambda) = - \epsilon_{n}(\lambda)\, .
\label{E23}
\end{equation}
In other words, the energy of the particles in TBA Eqs.(\ref{E14})
corresponds to the energy of the holes in BA Eqs.(\ref{E20}) of the Kondo
problem. The $\alpha$-type particle in the TBA equations is made
corresponding to the $n$-th string solution of Eqs.(\ref{E20}).
By changing the index $n$ by $\alpha$ we can
rewrite Eqs.(\ref{E20}) in the new notations as follows
\begin{eqnarray}
\epsilon_{\alpha}^{f}(\lambda) &=& \frac{2\epsilon_{F}}{\pi}\tan^{-1}\exp (\pi
\lambda)\delta_{\alpha 1} - \frac{T}{2\pi}\int_{-\infty}^{\infty}
\frac{1}{\cosh (\lambda-\lambda')} \nonumber \\
& &\left[\ln\left(1+e^{-\epsilon_{\alpha -1}^{f}(\lambda')/T}\right)+
\ln\left(1+e^{-\epsilon_{\alpha +1}^{f}(\lambda')/T}\right)\right]d\lambda'\, .
\label{E24}
\end{eqnarray}
The boundary conditions have now the form $\epsilon_{0}^{f}=\infty$
and $\lim\limits_{\alpha\to\infty}\epsilon_{\alpha}^{f}/\alpha=-H$.
From Eqs.(\ref{E24}) and Eqs.(\ref{E14}) we have
\begin{equation}
\Phi^{f}(\lambda-\lambda') =  \frac{1}{\cosh (\lambda-\lambda')}
\cal L
\label{E25}
\end{equation}
and
\begin{equation}
\epsilon_{\alpha}^{0}(\lambda) = \frac{2\epsilon_{F}}{\pi}\tan^{-1}\exp (\pi
\lambda)\delta_{\alpha 1} \, ,
\label{E26}
\end{equation}
where
\begin{equation}
{\cal L}=
\left(
\matrix { 0    & 1      & \null  &
                 \smash{\lower1.7ex\hbox{\LARGE 0}} \cr
         1     & 0  & \ddots & \null   \cr
        \null  & \ddots  &  \ddots     &  1      \cr
     \smash{\hbox{\LARGE 0}}   & \null   &  1    & 0  \cr}
\right) \, .
\label{E27}
\end{equation}
Unlike in the TBA Eqs.(\ref{E14}), we know the function 
$\epsilon_{\alpha}^{0}
(\lambda)$ in Eqs.(\ref{E26}) because it was found by the BA method.
\par
Let us assume  now that the particles have the GES with any statistical
matrix $\cal G$. In this case the matrix $\Phi$ in Eqs.(\ref{E15})
for the Kondo problem is
\begin{equation}
\Phi(\lambda-\lambda') =  \frac{1}{\cosh (\lambda-\lambda')}
{\cal L} - 2\pi\delta(\lambda-\lambda')\left[\cal I-\cal G\right]\, .
\label{E28}
\end{equation}
It is easy to see from the last expression that in the general case
we cannot find a matrix $\cal G$ to set the matrix $\Phi$ to zero. 
However,
it is possible  to consider the particular cases of high- and 
low-temperature
limits. The asymptotic behavior of the spin free energy
determines the different values of the rapidity range required
for the consideration of the high- and low-temperature limits.
When $T\to\infty$ (weak coupling limit), the impurity free energy
is given by
\begin{equation}
F_{imp} = -\frac{T}{2}\ln\left( 1+\exp \left(-\frac{\epsilon_{2S}^{f}
(-\infty)}
{T}\right)\right)\, .
\label{E29}
\end{equation}
When $T\to 0$ (strong coupling limit),
\begin{equation}
F_{imp} =  -\frac{T}{2}\ln\left( 1+\exp\left(-\frac{\epsilon_{2S}^{f}
(+\infty)}
{T}\right)\right)\, .
\label{E30}
\end{equation}
Therefore, in these temperature limits we have to consider 
$\epsilon_{2S}^
{f}(\lambda)$ at $\vert\lambda\vert\to\infty$. Taking the limit
$\vert\lambda\vert\to\infty$ in Eqs.(\ref{E24}) we have the 
equations
for $\epsilon_{\alpha}^{f}(\pm\infty)$:
\begin{equation}
\epsilon_{\alpha}^{f}(\pm\infty) =  -\frac{T}{2}\ln\left[\left(1+
\exp \left( -\frac{\epsilon_{\alpha -1}^{f}(\pm\infty)}{T}
\right)\right)
\left(1+\exp\left( -\frac{\epsilon_{\alpha +1}^{f}(\pm\infty)}
{T}\right)\right)
\right]\, .
\label{E31}
\end{equation}
The boundary conditions are: $\lim\limits_{\alpha\to\infty}
\epsilon_{\alpha}^{f}
(\pm\infty)/\alpha =-H$, $\epsilon_{0}^{f}(-\infty)=\infty$ 
and $\epsilon_{1}^
{f}(+\infty)=\infty$.
The solution of these equations is \cite{tz3}
\begin{equation}
\epsilon_{\alpha}^{f}(+\infty) =  \epsilon_{\alpha +1}^{f}(-\infty) =
-T\ln\left[\left(\frac{\sinh\left[ \frac{H}{2T}(\alpha +1)\right] }
{\sinh (\frac{H}{2T})}\right)^{2}-1\right]\, .
\label{E32}
\end{equation}
\par
This result will be the same if we assume that the matrix
$\Phi^{f}(\lambda)$ in Eqs.(\ref{E24}) is proportional to the
$\delta$-function, i.e.,
\begin{equation}
\lim_{\vert\lambda\vert\to\infty}\Phi^{f}(\lambda-\lambda') =
\pi\delta(\lambda-\lambda')\cal L \, .
\label{E33}
\end{equation}
It is clear because the function $\lim\limits_{\lambda\to\infty}1/
[\cosh (\lambda-\lambda')]=2/[\exp (\infty-\lambda')]$ acts as the
$\delta$-function at $\lambda'=\infty$.
Therefore, in the GES case the matrix $\Phi(\lambda)$  has
the following form
\begin{equation}
\Phi(\lambda-\lambda') = \pi\delta(\lambda-\lambda')
\left[{\cal L} - 2 {\cal I} + 2 {\cal G}\right]
\label{E34}
\end{equation}
for these temperature limits.
From the condition $\Phi=0$ of the ideal statistics we have the
following form of the statistical matrix:
\begin{equation}
{\cal G}=\frac{1}{2}
\left(
\matrix {2     & -1      & \null  &
                 \smash{\lower1.7ex\hbox{\LARGE 0}} \cr
        -1     &  2  & \ddots & \null   \cr
        \null  & \ddots  &  \ddots     & -1      \cr
        \smash{\hbox{\LARGE 0}}   & \null   &  -1    & 2  \cr}
\right) \, .
\label{E35}
\end{equation}
We see that it is proportional to the Cartan matrix for algebra, 
$A_{n}$.
\par
To find the distribution function $n_{\alpha}$ we need to know
the structure of the low-lying excitations. It follows from
Eq.(\ref{E24}) that $\epsilon_{\alpha}(\lambda)=0$ for all 
$\alpha\ne 0$ when
$\lambda=-\infty$ and for all $\alpha\ne 1$ when $\lambda=+\infty$.
The structure of the equations for the distribution function of
particles $n_{\alpha}$, which can be obtained from Eq.(\ref{E6})
with the statistical matrix (\ref{E35}), resembles that of the 
equations
for $\epsilon_{\alpha}(\vert\lambda\vert = \infty)$ (\ref{E31}):
\begin{equation}
\left( \frac{1}{n_{\alpha}}-1\right)^2 =
\left( 1- n_{\alpha+1}\right)\left( 1-n_{\alpha-1}
\right)\, .
\label{E36}
\end{equation}
The solution of Eqs.(\ref{E36}) (with the boundary condition
$\lim\limits_{\alpha\to\infty}\left(\frac{1}{n_{\alpha}}-1\right)
\to\exp (-\alpha H/T)$)
coincides with the solutions (\ref{E32}) :
\begin{equation}
\left( \frac{1}{n_{\alpha}}-1\right) =
e^{\epsilon_{\alpha}^{f}/T} =
\left[\left(\frac{\sinh (\frac{H}{2T}(\alpha+1)}{\sinh (\frac{H}{2T})}
\right)^{2}-1\right]
\label{E37}
\end{equation}
at $T\to 0$ and
\begin{equation}
\left( \frac{1}{n_{\alpha}}-1\right) =
e^{\epsilon_{\alpha}^{f}/T} =
\left[\left(\frac{\sinh (\frac{H}{2T}\alpha )}{\sinh (\frac{H}{2T})}
\right)^{2}-1\right]
\label{E38}
\end{equation}
at $T\to\infty$.
The spin free energy does not vary when we change the particle
statistics.
\par
The function $\Phi^{f}$ does not change in the TBA equations
for multichannel Kondo effect \cite{noz,ts1,ts2}. But the function
$\epsilon_{\alpha}^{0}(\lambda)$ is changed. It depends on the number
of channels, $k$, as
\begin{equation}
\epsilon_{\alpha}^{0} = \delta_{\alpha k}\exp{\pi\lambda}\, .
\label{E39}
\end{equation}
This leads to the new solutions for $\epsilon_{n}^{f}(\lambda)$
in the low-temperature limit $(~\lambda~\to ~+~\infty~)$:
\begin{eqnarray}
\epsilon_{\alpha}^{f}(+\infty) = \cases{
-T\ln\left[\left(\frac{\sin (\pi (\alpha +1)/(k+2)}{\sin (\pi /(k+2))}
\right)^{2}-1\right] \qquad \alpha <k \cr
-T\ln\left[\left(\frac{\sinh (\frac{H}{2T}(\alpha +1-k)}{\sinh (\frac{H}{2T})}
\right)^{2}-1\right] \qquad \alpha\geq k\cr }\, .
\label{E40}
\end{eqnarray}
The solutions for the higher-temperature limit $(\lambda\to -\infty)$
are not changed. To find the solution (\ref{E40}) using the
statistical matrix (\ref{E35}) we have to impose the additional
boundary condition $n_{\alpha}=1$ at $\alpha=k$.
\section*{Conclusion}
\par
Let us discuss in conclusion what new insights into the
thermodynamics of integrable systems (the Kondo problem
in particular) are gained by considering these systems
in terms of the GES.
\par
Using the GES principle we introduce an additional "parameter",
that is, particle statistics determined by the form of the
statistical matrix. If we suppose that the statistical matrix is
arbitrary, we can write the TBA equations for a system of
particles with any statistics. Each form of the statistical matrix
in this case has a corresponding distribution- and the 
DPS-function.
Therefore, we can write the TBA equations in a more convenient 
form.
A successful choice of the statistical matrix may lead to a 
simpler form
of the coupled equations for the dressed energy and for 
the distribution
function. In particular, in the case of ideal statistics
one can find a statistical matrix such that the solution of
the TBA equations for the dressed energy coincides with the bare 
energies,
and the equations for the distribution function repeat
(up to transformation) the TBA equations for Fermi particles.
\par
Let us focus on the features of the Kondo problem which are
studied using the GES principle. The consideration of spin 
excitations
as quasiparticles in the Kondo problem leads to the conclusion 
that
spin excitations correspond to the holes in the TBA approach.
To find the distribution function in the system of particles 
with
internal degrees of freedom we need additional information 
besides the
statistical matrix. We should know the detailed structures 
of the ground state
\cite{fu}. The energy of the ground state shows itself as the 
boundary
conditions for the equations determining the distribution 
function.
The boundary conditions include the external magnetic field which
compensates the internal "conformal" magnetic field of the system.
\par
The solution for the distribution function in 
high and low-temperature
regions has the universal form. It is determined by the
$q$-deformed dimension $[~\alpha~+~1~]~_{q}$ of irreducible 
representations of
the quantum group $U_{q}(sl_{2})$. 
Here $[~x~]_{~q~}~=~(q^{x}-q^{-x})/(q-q^{-1})$.
In the case of the single channel Kondo problem, the deformation
parameter $q~=~\exp ~(~H~/~2~T~)$ is real. It depends on the external
magnetic field and the temperature. In the multichannel Kondo problem
(at $\alpha < k$), the deformation parameter $q= \exp [i\pi /(k+2)]$
being the root of unity, is determined by the number, $k$ of the
channels.
\par
In summary, we used the TBA equations for a system of particles with
internal degrees of freedom to consider the features of the GES in the
Kondo problem. It is shown that the statistical matrix and the
distribution function in the Kondo problem have a universal form in 
high- and low-temperature limits.

\section*{Acknowledgments}
This work was supported in part by the Russian Foundation
for Basic Research under Grants Nos. 95-02-05620 and 
96-02-19272.


\newpage

\begin{thebibliography}{99}
\bibitem{hal} F. D. M. Haldane, {\it Phys. Rev. Lett.}
{\bf 67}, 937 (1991).
\bibitem{wu} Y.-S. Wu, {\it Phys. Rev. Lett.}
{\bf 73}, 922 (1994).
\bibitem{na} C. Nayak and F. Wilczek, {\it Phys. Rev. Lett.}
{\bf 73}, 2740 (1994).
\bibitem{isak} S. B. Isakov, {\it Phys. Rev. Lett.}
{\bf 73} 2150 (1994); {\it Int. J. Mod. Phys.} {\bf A9}, 2563 (1994).
\bibitem{pa} A. K. Pajagopal, {\it Phys. Rev. Lett.} 
{\bf 74}, 1048 (1995).
\bibitem{mur} M. V. N. Murthy and R. Shankar, {\it Phys. Rev. Lett.}
{\bf 72}, 3629 (1994).
\bibitem{prot} A. Protogenov, "{\it Haldanes statistical interactions
and universal properties of anyon systems}", IC/95/26 preprint, March 1995.
\bibitem{fu} T. Fukui and N. Kawakami, {\it Phys. Rev.}
{\bf B51}, 5239 (1995); {\it J. Phys.} {\bf A28} 6027.
\bibitem{cal} F. Calogero, {\it J. Math. Phys.}
{\bf 10}, 2197 (1962).
\bibitem{suth} B. Sutherland, {\it J. Math. Phys.}
{\bf 12}, 246 (1971); {\it Phys.Rev.} {\bf A4}, 2019 (1971).
\bibitem{ha} Z. N. C. Ha, {\it Phys. Rev. Lett.}
{\bf 73}, 1574 (1994).
\bibitem{ber} D. Bernard and Y.-S.Wu, in {\it New developments
of integrable systems and long-ranged interaction models,} ed.
M. L. Ge and Y.-S. Wu (World Scientific, Singapore, 1995).
\bibitem{wad} M. Wadati, {\it J. Phys. Soc. Jpn.}
{\bf 64}, 1552 (1995).
\bibitem{yang} C. N. Yang and C. P. Yang, {\it J. Math. Phys.}
{\bf 10}, 1115 (1969).
\bibitem{wen} X. G. Wen and A. Zee, {\it Phys. Rev.}
{\bf B46}, 2299 (1992).
\bibitem{noz} P. Nozi\`eres and A. Blandin, {\it J. Physique}
{\bf 41}, 193 (1980)
\bibitem{zam} Al. B. Zamolodchikov, {\it Nucl. Phys.}
{\bf 342}, 695 (1991).
\bibitem{tz3} A. M. Tsvelik and P. B. Wiegmann, {\it Adv. Phys.}
{\bf 32}, 453 (1983).
\bibitem{ts1} A. M. Tsvelik and P. B. Wiegmann, {\it J. Phys.}
{\bf A17}, 2321 (1983).
\bibitem{ts2} A. M. Tsvelik, {\it J. Phys.}
{\bf C18}, 159 (1985).
\end{thebibliography}
\end{document}